\documentclass[12pt]{iopart}
\usepackage{url} 
\usepackage{graphicx}
\usepackage{wrapfig}
\usepackage{harvard}
\begin{document}

\title[Radiation therapy calculations using a cloud computing cluster]{Radiation therapy calculations using an on-demand virtual cluster via cloud computing}

\author{R W Keyes$^1$, C Romano$^2$,
 D Arnold$^2$ and S Luan$^2$}
\address{$^1$ Department of Physics and Astronomy, University of New Mexico, Albuquerque, NM, 87131, USA}
\address{$^2$ Department of Computer Science, University of New Mexico, Albuquerque, NM, 87131, USA}
\ead{roy@unm.edu}

\begin{abstract}
Computer hardware costs are the limiting factor in producing highly accurate radiation dose calculations on convenient time scales. Because of this, large-scale, full Monte Carlo simulations and other resource intensive algorithms are often considered infeasible for clinical settings. The emerging cloud computing paradigm promises to fundamentally alter the economics of such calculations by providing relatively cheap, on-demand, pay-as-you-go computing resources over the Internet. We believe that cloud computing will usher in a new era, in which very large scale calculations will be routinely performed by clinics and researchers using cloud-based resources. In this research, several proof-of-concept radiation therapy calculations were successfully performed on a cloud-based virtual Monte Carlo cluster. Performance evaluations were made of a distributed processing framework developed specifically for this project. The expected 1/n performance was observed with some caveats. The economics of cloud-based virtual computing clusters versus traditional in-house hardware is also discussed. For most situations, cloud computing can provide a substantial cost savings for distributed calculations.
\end{abstract}




\maketitle

\section{Introduction}

Dosimetric calculations are a crucial component in the precision delivery and assessment of both radiation therapy and medical imaging. While Monte Carlo techniques are widely seen as the gold standard of radiation calculations, they are only sparingly used clinically, in favour of faster, less resource intensive algorithms at the cost of dosimetric accuracy \cite{rogers,tg105}. The primary barrier to widespread adoption of Monte Carlo techniques has been the requirement of large computing resources to achieve clinically relevant run times. These resources, usually in the form of a computing cluster, require a sizeable infrastructure investment as well as associated utility, maintenance, upgrade, and personnel costs. These costs make full Monte Carlo methods effectively infeasible for routine clinical use. It appears that the emerging Cloud Computing paradigm will remove this barrier, by making the necessary resources widely available in an economical way.

\begin{figure}[!h]
    \centering
		\includegraphics[width=0.9\textwidth]{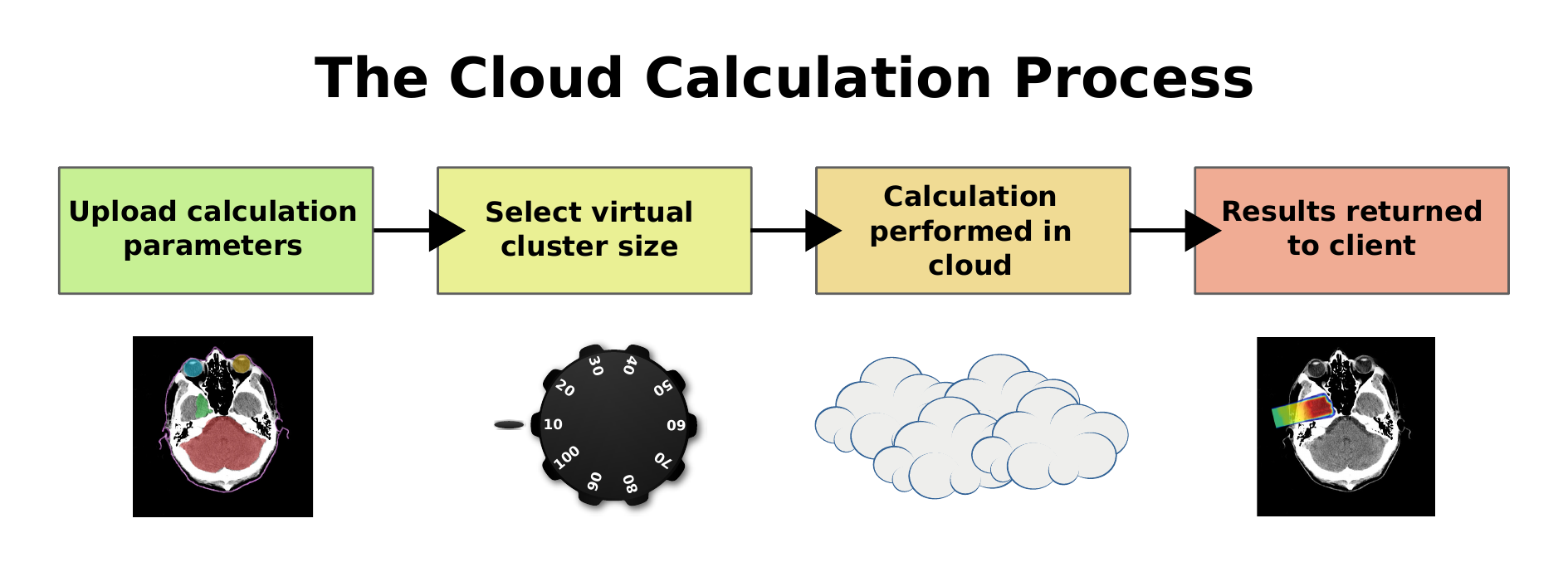} 
	\caption{The cloud calculation process as seen by the end user.}
	\label{fig:process}
\end{figure}

\subsection{CPU intensive calculations in radiation therapy}
As radiation therapy and diagnostic imaging techniques have become more complex, the associated physics calculations have become more resource intensive. These requirements have been largely met by the exponential increase in processor speed and RAM size, but sometimes outstrip the pace of computer technology even for conventional, deterministic calculation techniques. For example, TomoTherapy, Inc.'s TomoHD ships with a 14 node calculation cluster \cite{tomo}. 

Non-deterministic algorithms, such as the Monte Carlo method, demand even greater computing resources than conventional algorithms, but generally offer superior dose calculation accuracy. This is particularly true for complex, heterogeneous treatment scenarios and particle therapy treatment planning, but Monte Carlo has not yet been put into routine clinical use due to long calculation times. For example, Paganetti et al. have reported times of more than 100 CPU hours to simulate proton beam treatment plans when using approximately $2\times 10^7$ primary protons per field \cite{paga}.

\subsection{Cloud computing}
Cloud computing is a name given to a set of technologies offered as services over the Internet \cite{cloud,foster}. Cloud service providers, such as Google Inc., Amazon Inc., and Microsoft Inc., offer online computing resources (i.e. CPU time, storage, software, etc.), which are scalable to the user's need. Pricing is usually based on a pay-­as-­you-­go model, generally billed in hourly increments, and without set contract periods. This scheme allows cloud services to offer on­-demand computing infrastructure, sized to fit the user's momentary needs. Cloud computing has become feasible because of the economies of scale afforded by the commoditization of computer hardware, extensive availability of high bandwidth networks, and growth of free, open source software, including entire operating systems, such as Linux, and virtual machine software.

        For clinical usage the cloud computing paradigm has many potential advantages. Cloud resources can be scaled to meet patient and physics QA demand as it fluctuates on a daily basis. Typical computing clusters often face ``bursty'' usage: left under-utilized much of the day (and night) and over-queued at peak periods. The cloud paradigm is particularly well suited for one­-off calculations, such as machine commissioning and vault shielding calculations, for which a very large cluster might be desirable, but expanding to even a small cluster would be prohibitively expensive for a single run. Additionally, hardware upgrade and maintenance is taken care of by the provider, rather than by the user.

        Monte Carlo calculations are well suited to cloud style distributed computation by their fundamental nature, as the primary particle histories are completely independent of one another, requiring no communication between processes. This means calculations, while parallel, need not maintain data or timing synchronisation during execution. The first successful attempts to perform scientific Monte Carlo calculations in a cloud environment were made in 2007 by researchers associated with the RHIC experiment at Brookhaven National Laboratory \cite{RHIC} in the United States. Again in 2009, high energy physicists made use of a commercial cloud service to simulate a detector, this time from the KEK experiment in Japan \cite{KEK}. The cloud computing model has not been applied to medical physics calculations until the present research.
        
        An approach similar to cloud computing, called grid computing, has been utilized previously to perform distributed Monte Carlo calculations for radiation therapy \cite{Gomez06,Gomez07,Downes09}. Grid computing can be seen in many ways as the forerunner to cloud computing, having developed many of the distributed computing technologies and ideas upon which cloud computing is based \cite{foster}. The fundamental differences between the grid and the cloud lie in the virtual machine abstraction of the cloud and commercial implementation of the cloud infrastructure. The virtualisation in the cloud, abstracts away the particulars of the network and hardware architecture, so that they are transparent to the end user. Cloud providers are estimated to offer a minimum of an order of magnitude more total resources than all grid providers. For example, according to Rick Rashid, Chief of Microsoft Research, just a few Internet companies (including Google, Inc., Yahoo, Inc., Amazon, Inc., and Microsoft, Inc.) currently purchase some 20\% of all server computers \cite{waters}. The primary advantage of the cloud over the grid is the ubiquitous access to the cloud and the significantly lower barriers to entry by users, positioning the cloud to play a significantly larger role in medical physics computing.
        
        To date, the cloud computing model has made inroads into health care computing, primarily in the role of storing and processing medical records, data, and diagnostic images \cite{philbin,Andriole,langer}. A number of commercial vendors have begun to offer cloud-based PACS and related radiology software and services \cite{pacsdrive,emix,seemyradiology,carestream,lifeimage,insiteone,candelis}. An important aspect of such software is handling patient data privacy and relevant security. All of the vendors make claims to be in compliance with applicable regulations, such as the Health Insurance Portability and Accountability Act (HIPPA) in the United States.
        
\begin{center}
\begin{table}
    \begin{tabular}{ | c | c | c | c | }
    \hline
    EC2 Instance &  Small & XL (High CPU) & Quad XL (High RAM) \\ \hline
    Compute Units & 1 & 20 & 26 \\ \hline
    USD/hr & 0.085 & 0.68 & 2.0 \\ \hline
    USD/(hr $\times$ Compute Unit) & 0.085 & 0.034 & 0.0769 \\ \hline
    Architecture & 32-bit & 64-bit & 64-bit \\ \hline
    Memory & 1.7GB & 7GB & 68.4GB \\ \hline
    Storage & 160GB & 1690GB & 1690GB \\ \hline
    \end{tabular}
	\caption{Amazon EC2 pricing for instances running Linux in US eastern region (as of April 2010). An EC2 Compute Unit is equivalent to a 1.0­-1.2 GHz 2007 Opteron or 2007 Xeon processor.}
	\label{ec2-price}
\end{table}
\end{center}

\section{Methods and materials}
To demonstrate the feasibility of performing medical physics calculations using the cloud computing paradigm, we performed several ``typical'' physics calculations, including photon, electron, and proton beam depth-dose curves and a simple proton beam treatment plan calculation with CT-derived voxel data. Our calculations were carried out on Amazon Inc.'s Elastic Compute Cloud (EC2) service \cite{ec2}. At the time of our research, the EC2 service had the largest userbase and the most mature application programming interface, and was thus chosen for the tests. Several other vendors offer similar cloud services and would have been appropriate for this research. Amazon's EC2 offers several different processor-­RAM combinations at different hourly rates (see Table \ref{ec2-price}). Each processor is rated in terms of \textit{EC2 Compute Units}, which Amazon claims is equivalent to a 1.0­-1.2 GHz 2007 Opteron or 2007 Xeon processor.

Our calculations were carried out on the default EC2 Small instances with 1.7 GB of RAM, 1 virtual core, 
160 GB of local disk storage, and a 32­-bit architecture. Each EC2 instance runs inside a virtual machine on Amazon's servers. An operating system with user configured software is loaded using an Amazon Machine Image (AMI). AMI's can be chosen from a pre-configured set provided by Amazon, found elsewhere online, or built by the user. From the user perspective, the boot-up of an instance using a pre-configured machine image is no different than starting a standard server computer. Once the instance is up and running, it will have a unique IP address and a domain name, allowing the user to log in. A virtual cluster can be built by simply requesting multiple virtual nodes (instances). The size of the cluster can be scaled on demand (i.e., virtual nodes can be dynamically created and destroyed). Files were stored on the the running EC2 instances and on the persistent Amazon Simple Storage Service (S3) \cite{S3} to facilitate transfer to and from the cloud. From a user perspective, the Amazon S3 storage service can be viewed as the counterpart of 
the underlying network file system (NFS) found in most cluster computing environments.

To perform distributed dose calculations and output processing, we implemented a custom distributed processing framework. This framework, dubbed \textit{flsshd}, used the Python programming language, the  boto Python library to access Amazon Web Services \cite{boto}, and the secure shell protcol (SSH) for network communication. We used an AMI built from Fedora Linux. Monte Carlo simulations were carried out with the Fluka Monte Carlo package (version 2008.3b) \cite{fluka1,fluka2}.

\begin{figure}[!h]
    \centering
		\includegraphics[width=0.8\textwidth]{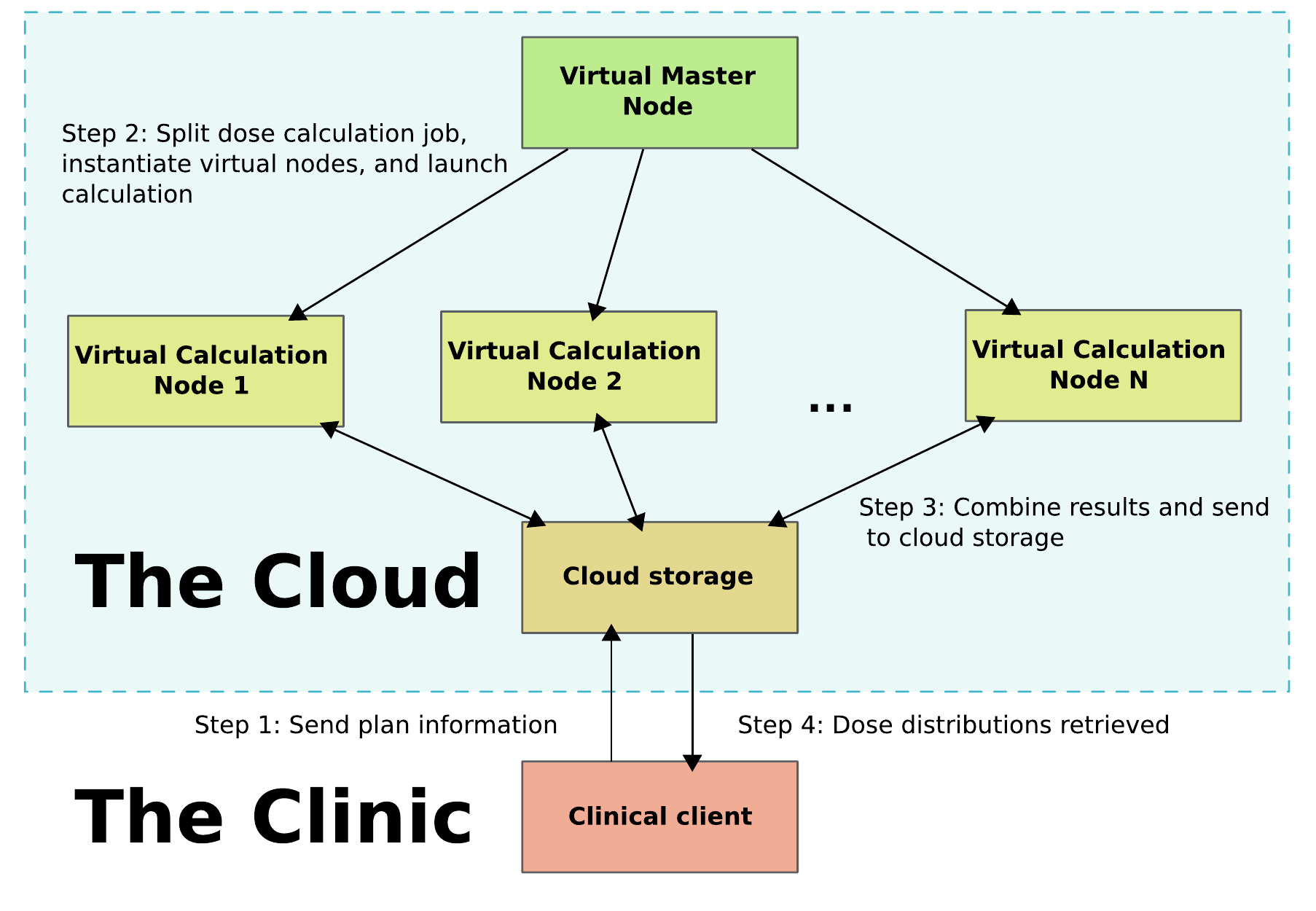} 
	\caption{General scheme for calculating dose using the cloud paradigm.}
	\label{fig:model}
\end{figure}

The general data flow of the cloud based calculations is as follows (see Fig. \ref{fig:process} and \ref{fig:model}): (1) the client computer uploads the calculation parameter (input) file to the online storage. (2) The client requests N nodes for the calculation. (3) The input file is distributed to each node, given a unique, random seed, and the Monte Carlo calculation is carried out. (4) Once the dose calculations are completed, the nodes collect the resulting dose files and combine them into a single output  file, which is returned to the client. The result combination step can occur on a single master node or in parallel.

Proof of concept studies included calculating depth-dose curves for a 75~MeV proton pencil beam, a $10 \times 10$~cm$^2$, 100~cm SSD Co-­60 beam, a $10 \times 10$~cm$^2$, 100~cm SSD 10~MeV electron beam, and a simple broad-­beam $3 \times 3$~cm$^2$, single angle proton plan using a voxel phantom of heterogeneous tissues based on CT data \cite{zubal}. All depth-dose calculations were performed in a virtual $40 \times 40 \times 40$~cm$^3$ water phantom. Cluster performance tests were carried out using the proton depth-­dose calculations with beams of 75~MeV and 200~MeV. To determine the dependence of total run time on number of virtual nodes, a proton simulation with a total of $1.4 \times 10^7$ primary protons was carried out on between 1 and 200 virtual nodes. The number of primaries was evenly divided between each node (i.e. each node in the 10 node calculation processed $1.4 \times 10^6$ primaries). Each data point (number of nodes) was run 3-4 times. The total run time was described using a simple model: 

\begin{equation}
T(n) = \frac{\alpha p}{n} + \beta n + \gamma
\label{eq:1}
\end{equation}
where $T$ is the total calculation time, $n$ is the number of virtual nodes, $p$ is the number of primary particles (or histories), and $\alpha$, $\beta$, and $\gamma$ are constants. $\alpha$ is interpreted as the calculation time per primary, $\beta$ is interpreted as time to initiate the calculation on each node that scales with $n$, and $\gamma$ is interpreted as the fixed overhead time (e.g. node startup and shutdown times). This simple model ignores possible higher order node communication effects.

\begin{figure}[!h]
    \centering
		\includegraphics[width=0.4\textwidth]{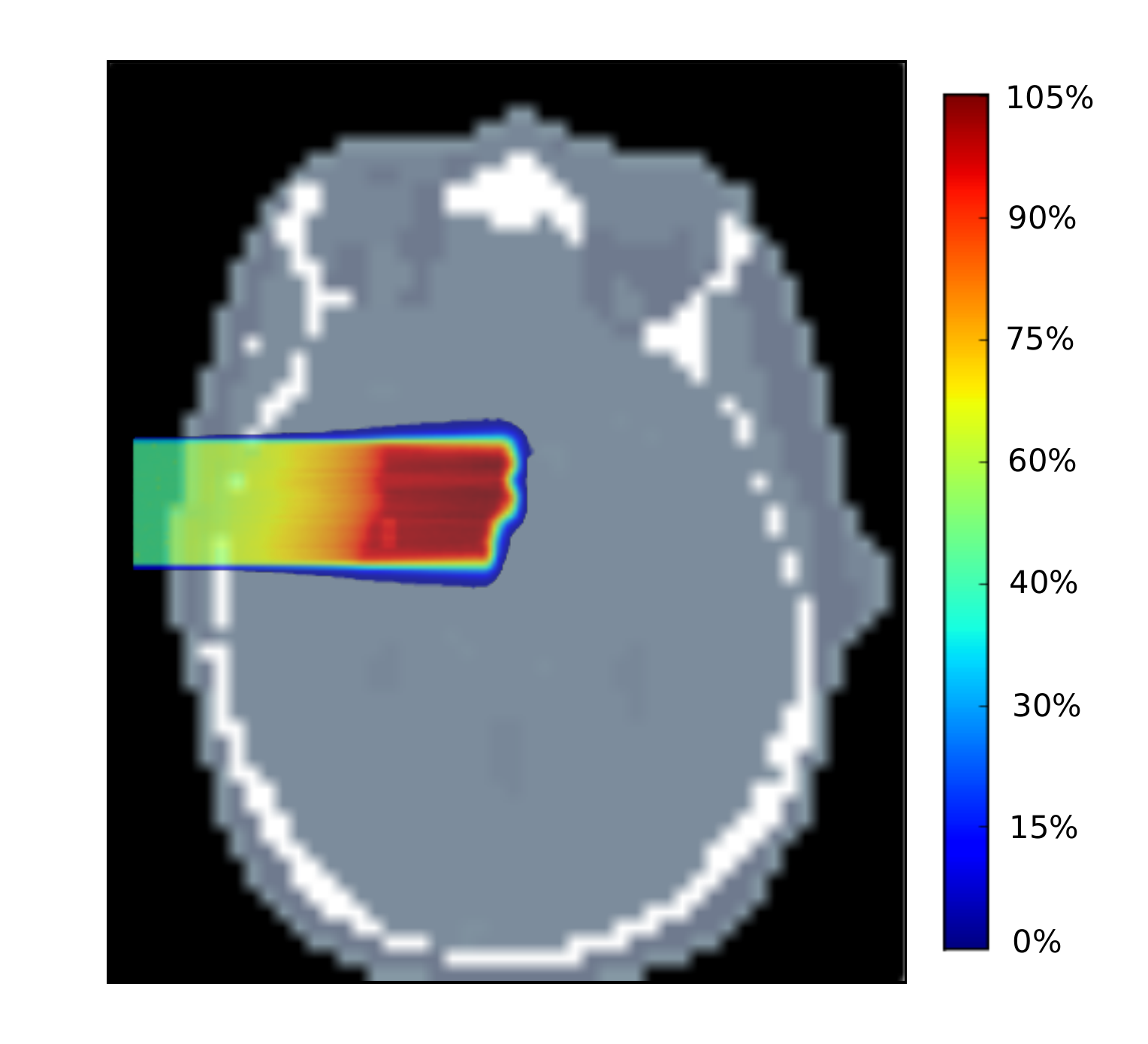} 
	\caption{A simple broad-beam proton therapy plan calculated on 130 virtual nodes on Amazon's EC2 cloud service using a voxel based phantom.}
	\label{fig:plan}
\end{figure}

\section{Results}
\begin{figure}[!h]
    \centering
		\includegraphics[width=0.8\textwidth]{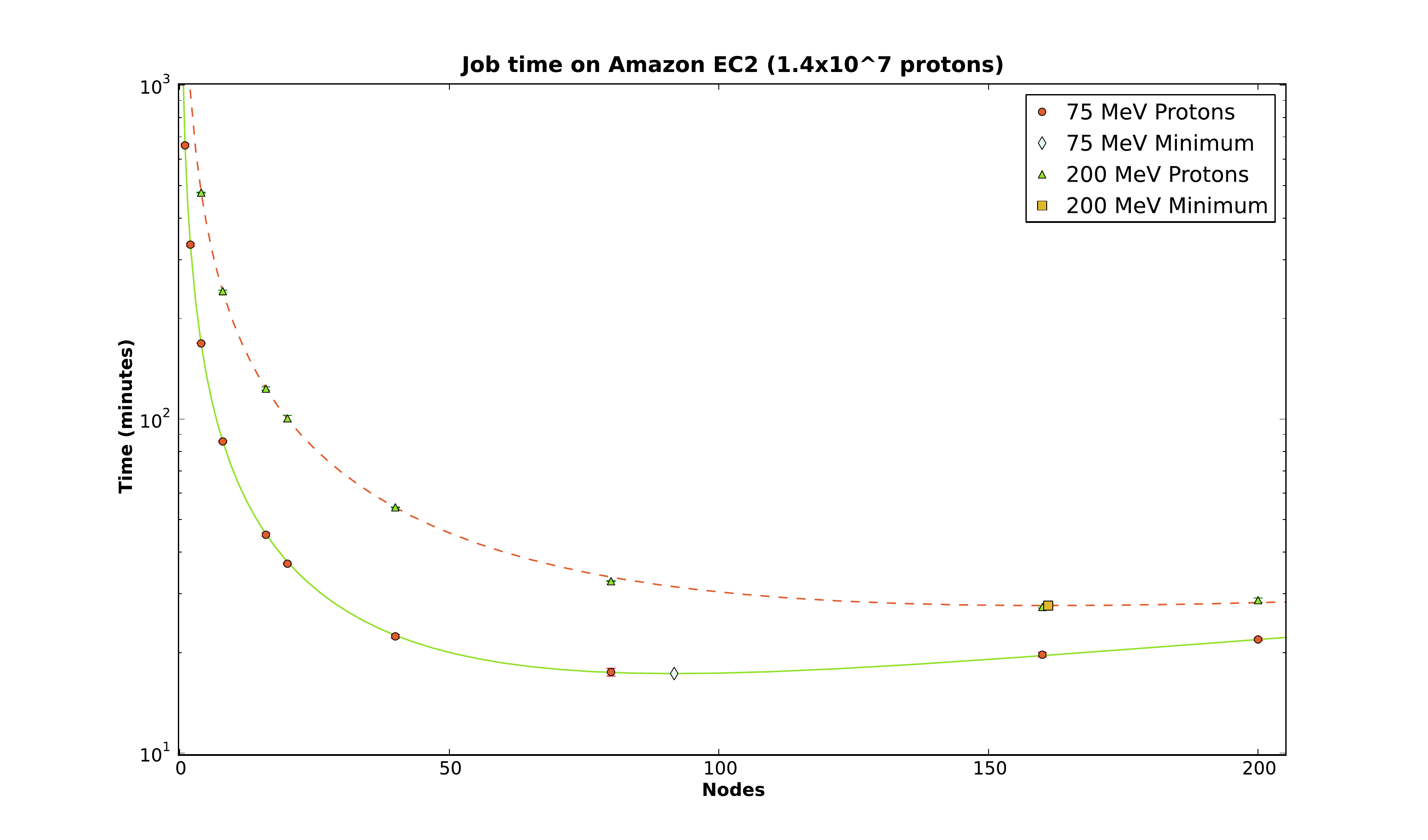} 
	\caption{Calculation time versus number of virtual nodes for proton depth-dose curves on Amazon's EC2 cloud service. The calculation time is modelled with Eq.\ref{eq:1}.}
	\label{fig:performance1}
\end{figure}

The flsshd framework was used to carry out the above described proof-of-concept calculations on a cloud-based, virtual cluster, running on between 1 and 200 virtual nodes. The simple proton therapy plan was calculated on 130 virtual nodes (see Fig. \ref{fig:plan}). The performance test data is shown on a semi-log plot in Figure \ref{fig:performance1}. The total calculation times as a function of number of nodes is plotted for proton depth-dose curves at two energies (75~MeV and 200~MeV). The run time model (Eq. \ref{eq:1}) was fit to the data sets. The R$^2$ test implied excellent fit, with R$^2 >$ 0.99999. 

\section{Discussion}
The flsshd distributed processing framework was successfully used to perform proof-of-concept radiation therapy calculations using a cloud-based, virtual cluster. The minimum possible run time currently possible with flsshd is limited by the linear overhead term (described by $\beta$ in Eq. \ref{eq:1}). This is due to the serial nature of how the simulations are initiated by flsshd, and is not a fundamental limitation in the performance of a cloud-based, virtual cluster. Some type of overhead will always be present, so a perfect 1/n speed-up will never be achievable for an arbitrarily large number of nodes. This fundamental asymptotic performance limit is described by Amdahl's Law applied to parallel processing \cite{hennessy}.

	Ultimately, the impetus behind using the cloud computing model as a computing cluster is the desire for large scale processing power without an associated large price tag. Costs associated with the cloud are generally only incurred on a usage basis, whereas in-house hardware incurs capital and maintenance costs. In order to compare the costs of the two models, we ignore the associated personnel, utility, equipment housing, insurance, vendor service, and other miscellaneous costs that might be associated with the in-house model. While this is unrealistic, it is very difficult to estimate the average costs of many of these categories (e.g. some departments may pay rent for their server space, while others need not). This set of assumptions also puts the in-house cluster in the best-case scenario.
	
	An informal survey of various equipment purchasers within the university estimated the cost of a computing cluster at approximately 1000 USD per node plus approximately 200 USD in maintenance costs per year per node. For a 100 node cluster, the approximate cost over the expected 3 year lifetime of the cluster would be approximately 160,000 USD (or 53k USD per year).
	
	The cost for using a cloud-based cluster is primarily determined by the number of CPU-hours used. We used the assumption that each patient needs approximately 100 CPU-hours of cluster time. This assumption was not based on \textit{typical} use of clinical Monte Carlo, because the types of calculations possible with an extremely large cluster are not typically performed in a clinical setting. Calculations which might be possible with ubiquitous access to extremely large clusters, such as treatment planning with iterative Monte Carlo optimization, are the subject of future research. Assuming that a typical patient throughput for a clinic is 1000 patients per year, the CPU time necessary would be be approximately 100,000 CPU-hours. Amazon's current cheapest offering allows for 1 EC2 Compute Unit to be purchased for 0.034~USD. This comes out to approximately 0.10~USD per 3~GHz CPU-hour, with the assumption that 3~GHz is approximately the speed of a contemporary high end CPU core. Thus the CPU costs come out to approximately 10,000 USD per year. Additional costs are added by storing data in and transferring data to and from the cloud. A conservative estimate of data transfer and storage for this application puts these costs at less than 5\% of the annual CPU costs. The resulting total annual cost using the AWS services was thus estimated at less than 10,500~USD. This was less than 20\% of the estimated annual in-house cluster cost.

	A major concern when transferring or storing personal or patient data on networked hardware is the security and privacy of the data. In many countries this is specifically regulated by the government. These regulations include
the {\it Health Insurance Portability and Accountability Act (HIPAA)} in the United States,
the {\it Personal Information Protection and Electronic Documents Act} in Canada,
{\it Information Technology, Files and Civil Liberties Act no. 78-17 of 6 January 1978} in France, and the {\it Data Protection Act} in the United Kingdom, among others. For the type of application investigated in this research, the technological aspects of securing the data are relatively well understood \cite{Stinson,Bishop02,Rhee,Bishop04}. Additionally, compared to full medical record or medical image storage, radiation treatment planning and similar calculations do not require as much personally identifiable information to be sent over the network (i.e. names can be replaced with hash tags). It is encouraging that there appears to be adoption of other similar cloud-based frameworks mentioned earlier that would require compliance with the applicable regulations. Cloud service provider Amazon has published its own set of guidelines to assist cloud developers in bringing their applications into compliance with the American HIPPA regulations covering patient data privacy \cite{awshippa}.
	
	Another technology that has gained widespread popularity in medical physics
computation in recent years is the so-called GPGPU \cite{Greef,Gu09,Men,Gu10}.
The GPU, which stands for {\it graphics processing unit}, is a specialized microprocessor that 
offloads and accelerates graphics rendering in a graphics card. The primary advantage of GPUs is their relatively large number of cores, offering parallel hardware at low cost.
GPGPU, which stands for {\it general propose computing on GPU}, is the technique of using a GPU 
to perform computation in applications other than graphics \cite{Owens}.
Many researchers have found the word ``general'' to be misleading, because
the GPU cannot seamlessly take on every task of a traditional 
CPU \cite{Owens}.
The arithmetic power of the GPU is a result of its highly specialized architecture. 
As a result, GPUs lack some fundamental computing constructs, such as integer 
data operands, and do not support 64-bit double precision arithmetic, among other deficits.
The GPU uses an unusual programming model \cite{PURCELL,BUCK,Pharr}, 
and effective GPU programming is not 
simply a matter of learning a new language, or writing a new compiler backend. For these reasons, the GPU architecture might not be appropriate or convenient for implementing certain algorithms.
In contrast, almost all of the current dose calculation engines and treatment planning
systems can be installed on cloud-based CPU clusters with little or no modification. The distributed part of the cloud calculation would be handled by external frameworks, such as flsshd, described here.

We believe cloud computing and the cost effective parallelism provided by GPGPU are complementary rather than competing technologies, each with its appropriate niches. Cloud computing is, in fact, a general model and several researchers and vendors have already proposed clouds with some mix of CPU and GPU resources to provide the best possible speed-ups  \cite{giunta,sgi,peer1}. Choosing the best distributed computing architectures and taking advantage of their respective strengths will play a key role in the future of medical physics computing. 

\section{Conclusion and Future work}
The emerging cloud computing paradigm appears to provide very interesting opportunities for computing in medical physics. We have successfully demonstrated a proof-of-concept distributed calculation framework, which utilizes an on-demand, virtual computing cluster run on a commercial cloud computing service. The on-demand nature, ease of access, and pay-as-you-go pricing model yield the promise of providing clinics and researchers access to unprecedented amounts of computing power for medical physics calculations in the coming years.

	Our custom calculation framework, flsshd, is currently designed to run with only one specific Monte Carlo engine, but demonstrates the conceptual basis for powerful distributed calculations using cloud computing, whether they are Monte Carlo or other resource intensive algorithms. For future work, we are: (1) building a web portal for researchers to upload their Monte Carlo calculations to the cloud, (2) converting flsshd into a fully distributed application, which includes fully parallel initiation of the simulations in the entire virtual cluster for increased performance, (3) extending our framework to other dose calculation engines such as Geant4 and EGS5 \cite{geant41,geant42,egs5}, and (4) extending commonly used optimization routines such as simulated annealing and constrained least squares solvers into the cloud. 
	
\ack
This work was supported in part by the National Science Foundation under
grant CBET-0853157. Usage of the AWS cloud services was partially covered by credits from an AWS in Education research grant from Amazon, Inc. 

\section{References}
\bibliography{pmb-cloud-paper.bib}

\end{document}